\documentstyle[aps,prb,floats,epsf]{revtex}
\newcommand{\smfrac}[2]{\mbox{\small $#1 \over #2$}}
\begin{document}
\draft
\twocolumn[\hsize\textwidth\columnwidth\hsize\csname@twocolumnfalse%
\endcsname
\title{A Common Universality Class for the Three--Dimensional
Vortex Glass and Chiral Glass?}
\author{Carsten~Wengel and A.~Peter~Young}
\address{Department of Physics, University of California, Santa
Cruz, California 95064}
\date{\today}
\maketitle
\begin{abstract}
We present a Monte Carlo study of the $d=3$ gauge glass
and the XY--spin glass models in the vortex representation.
We investigate the critical behavior
of these models by a scaling analysis of the linear resistivity and
current--voltage characteristics, both in the limits of zero
and strong
screening of the vortex--interactions. Without screening, both
models show a glass transition at a finite temperature
and, within the numerical accuracy, exhibit the {\em same} critical
exponents: $z\approx3.1$ and $\nu=1.3\pm 0.3$. With strong screening,
the finite temperature glass transition is destroyed in both cases
and the same exponent $\nu=1.05\pm 0.1$ is found at the resulting zero
temperature transition.
\end{abstract}
\pacs{PACS numbers: 7550.Lk, 7540.Mg, 7460.-w, 0550.+q}
]

\section{Introduction}

It has been suggested\cite{fisher89,ffh91} that defects may collectively pin
flux lines (vortices) in a type--II superconductor in a field, leading to a
vortex glass phase with vanishing linear resistance. In many numerical studies
of the vortex glass transition,
a simple model called the ``gauge glass'' has been used.
\cite{fisher91,reger91,hyman95,bokil95,wengel96}
A related model is the XY--spin glass, which has been studied extensively
in order to understand the magnetic ordering of a variety of magnetic
compounds with random and frustrated interactions.
XY--spin glasses are of special interest since they
potentially exhibit two distinct kinds of ordering: spin glass ordering due
to freezing of the spins, and ``chiral glass'' ordering due to freezing of
local chiral (vortex) degrees of freedom.
\cite{villain77,kawamura85,huse90,ray92,kawamura91,kawamura92,kawamura95/6,bokil96,kawamura97}
It has been well established\cite{jain86,jain96}
that the spins do not have a finite
temperature spin glass transition in three dimensions, whereas
Kawamura\cite{kawamura95/6,kawamura97} has argued that a finite temperature
chiral glass transition does occur. This intriguing claim provides one of the
main motivations for the present study.

In this paper we present a comprehensive Monte Carlo study of the vortex
glass transition in the gauge glass model and
the chiral glass transition in the XY--spin glass in three dimensions.
We consider both the situation where screening between the vortices is
neglected, (which is the case in most of the earlier work) and also where there
is strong screening of the vortices. We find that both with and without
screening, the chiral glass and
gauge glass have very similar behavior. Without screening they have
a finite temperature transition with
numerically very similar values for exponents,
suggesting that they may lie in the same universality class. 
For both models, we find that screening destroys the finite temperature
transition. 

Our paper is organized as follows: In Sec.~\ref{model}
we define the models under consideration.
In Sec.~\ref{method} we discuss the quantities that we calculate
and explain the finite size scaling techniques 
used in the analysis.
In Sec.~\ref{gg} we present our results for the gauge glass model
without screening, and briefly review results for the gauge glass
with screening that we found earlier in  Ref.~[\onlinecite{wengel96}]
(referred to as WY).
In Sec.~\ref{xy} we present
results for the the XY--spin glass with and without screening.
We summarize
our results and draw our conclusions in Sec.~\ref{conclude}.

\section{The Models} \label{model}

In the absence of screening the Hamiltonian of
both the XY--spin glass and the gauge glass
can be written in the {\em phase representation} as
\begin{equation}
\label{glass_oo}
{\cal H} = -J\sum_{
\langle i,j\rangle } \cos(\phi_i - \phi_j - A_{ij}),
\end{equation}
where the $\phi_i$ are interpreted either as phases of a superconducting
order parameter (gauge glass) or as the angles of two--dimensional spins
(spin glass). Here,
$J$ is the interaction strength (henceforth set to unity), and 
the sum is taken over all nearest neighbor sites $\langle i,j\rangle$ on
a simple cubic lattice. In the case of the gauge glass, the effects of
the external magnetic field and the disorder are represented by quenched
vector potentials $A_{ij}$, taken to be uniformly distributed in
the interval $[0,2\pi ]$. In the case of the $\pm J$ XY--spin glass,
the random sign of the bonds between spins is represented by quenched
vector potentials $A_{ij}$ taken randomly to be 0 (+$J$) or $\pi$
(-$J$).

The Hamiltonian (\ref{glass_oo}) obviously possesses a $U(1)$
symmetry, i.e. the model is invariant under the transformation
$\phi_i \to \phi_i + C\ \ \forall i$, where $C$ is a constant.
For the gauge glass this is the only symmetry. However, for the XY--spin glass
there is an additional ``reflection'' symmetry, $\phi_i \to -\phi_i
\ \ \forall i$.

It is convenient to rewrite the Hamiltonian in such a way that the chiral
(vortex) variables, which are our main concern, appear explicitly.
This transformation involves replacing the
cosine in Eq.~(\ref{glass_oo}) with the periodic 
Gaussian Villain function,
separating spin wave and vortex variables, and then 
performing fairly standard
manipulations\cite{jose77,dasgupta81,kleinert} to obtain 
\begin{equation} \label{vortexglass}
{\cal H}_V=-\frac{1}{2}\sum_{i,j} G(i-j)
\, [{\bf n}_i-{\bf b}_i] \cdot
[{\bf n}_j-{\bf b}_j].
\label{hv}
\end{equation}
Here, the vortex variables ${\bf n}_i \in \{0,\pm 1,\pm 2, ...\}$
sit on the links of the {\em dual}
lattice (which is also a simple cubic lattice here),
$G(i-j)$ is
the lattice Green's function
\begin{equation}
G(i-j)= {(2\pi)^2 \over L^3} \sum_{{\bf k}\neq 0}
\frac{1-\exp[{\rm i}\;{\bf k}\cdot ({\bf r}_i - {\bf r}_j)]}
{2\sum_{n=1}^d [1-\cos(k_n)] },
\label{gij}
\end{equation}
(with $d=3$),
and
the ${\bf b}_i$ are quenched fluxes given by $(1/2\pi)$ times the directed sum 
of the quenched vector potential $A_{ij}$ on the original lattice
surrounding the link on the dual lattice on which ${\bf b}_i$ lies.
Due to periodic boundary conditions, we have
the global constraints $\sum_i {\bf b}_i= \sum_i {\bf n}_i = 0$.
There are also the {\em local} constraints, 
$[\nabla\cdot {\bf n}]_i = [\nabla\cdot {\bf b}]_i = 0$, 
where the latter just follows trivially from the definition of ${\bf
b}_i$ as a lattice curl.

Since the Hamiltonian only depends on ${\bf n}_i-{\bf b}_i$ it is convenient to
discuss the distribution of the quenched fluxes
when all the weight is shifted into
the interval\cite{shift}
$0 \le b^\alpha_i < 1$ (where $\alpha$ is a Cartesian component).
For the gauge glass model, the distribution of the shifted 
$b^\alpha$ is uniform, i.e.
\begin{eqnarray}
P(b^\alpha)  & = & 1    \qquad (0 \le b^\alpha < 1) \nonumber \\
& = & 0 \qquad\mbox(\rm otherwise) ,
\label{dist-gg}
\end{eqnarray}
while for the $\pm J$ spin glass the shifted $b^\alpha$ have a bi-modal
distribution with equal weight at 0 (corresponding to an unfrustrated square on
the original lattice) and 1/2 (corresponding to a frustrated square):
\begin{equation}
P(b^\alpha) = {1 \over 2} \Bigl( \delta(b^\alpha) +
\delta(b^\alpha-\smfrac{1}{2}) \Bigr)  .
\label{dist-sg}
\end{equation}


Recent work on the gauge glass model and
on the XY--spin glass have investigated the role of
screening of the vortex--vortex interactions,
which is a relevant perturbation near the critical
temperature\cite{ffh91,bokil95}.
It was found by Wengel and Young (WY)\cite{wengel96} and
Ref.~[\onlinecite{bokil95}], that the vortex glass phase vanishes when
strong screening is included in the $d=3$ gauge glass model,
and subsequent work by Kawamura and Li\cite{kawamura97}
found the same effect for the chiral glass transition. 
We therefore also discuss the effects of screening here.

In the vortex representation, the Hamiltonian is still represented by
Eq.~(\ref{vortexglass}) but now the interaction $G(i-j)$ has the screened form
\begin{equation}
G(i-j)= {(2\pi)^2 \over L^3} \sum_{{\bf k}\neq 0}
\frac{1-\exp[{\rm i}\;{\bf k}\cdot ({\bf r}_i - {\bf r}_j)]}
{2\sum_{n=1}^d [1-\cos(k_n)] + \lambda_0^{-2} },
\label{gij_screen}
\end{equation}
where $\lambda_0$ is a bare screening length. Note that in the long wavelength
limit, the denominator is just $k^2 + \lambda_0^{-2}$.

In the simulations presented here, we consider just two cases:
(i) $\lambda = \infty$, where there is no screening and the interactions
between the vortices are long range, and (ii) $\lambda \to 0$,
where there is strong screening.
In the latter case
$G(r \ne 0) =(2\pi\lambda_0)^2$ with corrections which are
exponentially small, i.~e., of order $\exp(-r/\lambda_0)$.
Because $\sum_i ({\bf n}_i - {\bf b}_i)= 0$ we can always
add a constant to $G(r)$ for all $r$ without affecting the results.
We therefore add $-(2\pi\lambda_0)^2 $, as a result of which
the only interaction is on--site, and then divide the interaction by
$(2\pi\lambda_0)^2 $ to have a well defined limit for $\lambda_0 \to 0$.
The resulting Hamiltonian
then has the very simple form
\begin{equation}
{\cal H}_V = \frac{1}{2}\sum_i ({\bf n}_i - {\bf b}_i)^2 \qquad
(\lambda_0 \to 0) .
\label{hv_screen}
\end{equation}
Note\cite{comment_z}, however, that
${\cal H}_V$ is not trivial because the local constraint
$[\nabla\cdot {\bf n}]_i = 0$ effectively generates interactions
between the ${\bf n}_i$.

To summarize, we study four models in this paper:
\begin{enumerate}
\item
The gauge glass in the absence of screening.  The Hamiltonian is given by
Eq.~(\ref{hv}) where the
$G(i-j)$ are given by Eq.~(\ref{gij}),
and the distribution of the fluxes (shifted\cite{shift}
into the interval from 0 to 1) is given by Eq.~(\ref{dist-gg}).
\item
The gauge glass with strong screening. The Hamiltonian is given by
Eq.~(\ref{hv_screen}) in which the distribution of the (shifted) fluxes
is given by Eq.~(\ref{dist-gg}).
\item
The chiral glass (i.e. vortex degrees of freedom in the XY--spin glass) in the
absence of screening. The Hamiltonian is given by Eq.~(\ref{hv}) where the
$G(i-j)$ are given by Eq.~(\ref{gij}),
and the distribution of the (shifted) fluxes is given by Eq.~(\ref{dist-sg}).
\item
The chiral glass with strong screening. The Hamiltonian is given by
Eq.~(\ref{hv_screen}) in which the distribution of the (shifted) fluxes 
is given by Eq.~(\ref{dist-sg}).
\end{enumerate}

\section{Data Analysis}
\label{method}

We simulate the Hamiltonians in Eq.~(\ref{hv}) and~(\ref{hv_screen})
on simple cubic lattices with $N=L^3$ sites where $4 \le L \le 12$.
Periodic boundary conditions are imposed. We start with configurations
with all ${\bf n}_i = 0$, which clearly satisfies the constraints, and a
Monte Carlo move consists of trying to create a loop of
four vortices around a square. This trial state is accepted  with
probability $1/(1+\exp(\beta \Delta E))$,
where $\Delta E$ is the change of energy and $\beta=1/T$.
Each time a loop is formed it generates a voltage
$\Delta Q = \pm 1$ perpendicular to it's plane, the sign
depending on the orientation of the loop. This leads
to a net voltage\cite{hyman95}
\begin{equation}
V(t)=\frac{h}{2e} I^V(t) \quad\mbox{with}\quad
I^V(t)=\frac{1}{L\Delta t}  \Delta Q(t),
\label{voltage}
\end{equation}
where $I^V$ is the vortex--current
and $t$ denotes Monte Carlo ``time'' incremented by
$\Delta t$ for each attempted Monte Carlo move.
We will work in units where $h/(2e)=1$, and we set $\Delta t=1/(3N)$
so that an attempt is made to create or destroy one vortex loop
per square in each direction, on average, per unit time.

The linear resistivity
can be calculated from the voltage fluctuations via the
Kubo formula\cite{young94}
\begin{equation}
\rho_{\rm lin}=\frac{1}{2T}\sum_{t=-\infty}^{\infty} \Delta t\;
\langle V(t)V(0)\rangle.
\end{equation}
Here, $\langle \cdots \rangle$ denotes the combined thermal and
disorder average.
Near a second order phase transition the linear resistivity obeys the
scaling law\cite{ffh91}
\begin{equation}
\rho_{\rm lin}(T,L)=L^{-(2-d+z)} \tilde{\rho}(L^{1/\nu}(T-T_c)),
\label{rho_lin_scale}
\end{equation}
where $\xi$ is the correlation length exponent, i.e.,
\begin{equation}
\xi \sim (T - T_c)^{-\nu} ,
\end{equation}
$z$ is the dynamical exponent,
and $\tilde{\rho}$ is a scaling function.
At the critical temperature, $\tilde{\rho}$ becomes a
constant and therefore $\rho_{\rm lin}(T_c,L) \sim L^{-(2-d+z)}$.
If we plot the ratio of $\rho_{\rm lin}$ for different system sizes
against $T$, then
\begin{equation} \label{intersect}
\frac{\ln[\rho_{\rm lin}(L)/\rho_{\rm lin}(L^{\prime})]}
{\ln[L/L^{\prime}]} = d-2-z \quad\mbox{at}\;\;  T_c,
\end{equation}
i.e., all curves for different pairs $(L,L^{\prime})$ should
intersect and one can read off the values of $T_c$ and $z$. We will
refer to this kind of data plot as the ``intersection method''.
With the values of $T_c$ and $z$ determined by the intersection method
we can then use a scaling plot according to Eq.~(\ref{rho_lin_scale})
to obtain the value of $\nu$.

In the case of strong screening we find a zero temperature transition
and a plot according to Eq.~(\ref{rho_lin_scale}) to determine $\nu$ 
does not work, since
$z=\infty$ because there is activated dynamical scaling at the $T=0$
transition. However, one can still obtain static exponents by measuring the
voltage generated by a finite external current, i.~e., by $I$--$V$
characteristics. In real superconductors, transport currents
generate a non-uniform magnetic field because of Amp\`ere's law,
$\vec \nabla \times {\bf B} = {\bf J}$. It is inconvenient to simulate a
non--uniform system, so instead we
effectively assume that the current is the same everywhere so
{\em each} vortex feels a Lorentz force ${\bf n}_i \times {\bf J}$.
The scaling behavior of the response to such a perturbation should be
the same as that derived earlier for response to an actual
transport current\cite{ffh91}. We can therefore use this approach to
determine critical exponents, which is our objective.
The Lorentz force biases the moves and sets up a net flow of vortices
perpendicular to the current, whose time average gives the voltage
according to\cite{hyman95} Eq.~(\ref{voltage}).

To analyze our data we need to understand the scaling behavior
of the $I$--$V$--curves near a second order phase transition. The scaling
theory gives\cite{ffh91,hyman95}
\begin{equation}
T\frac{E}{J}\frac{\tau}{\xi^{d-2}}=g\left(\frac{J\xi^{d-1}}{T}\right),
\label{iv_scale}
\end{equation}
where $E$ is the electric field, $J$ the current density, $\tau$ a relaxation
time, and
$g$ is a scaling function.
At a zero temperature transition one has
\begin{equation}
\xi \sim T^{-\nu},
\end{equation}
so, in three dimensions,
Eq.~(\ref{iv_scale}) becomes
\begin{equation}
T^{1+\nu} \frac{E}{J} \tau =g\left(\frac{J}{T^{1+2\nu}}\right).
\end{equation}
From this equation we can see that the current scale, $J_{\rm NL}$,
at which nonlinear
behavior sets in varies with $T$ as $J_{\rm NL}\sim T^{1+2\nu}$.
Since the linear resistivity is defined by
\begin{equation}
\rho_{\rm lin}=\lim_{J\to 0}\frac{E}{J},
\end{equation}
and $g(0)$ can be taken to be unity, we can write
\begin{equation}
\label{scale_iv}
\frac{E}{J\rho_{\rm lin}}=g\left(\frac{J}{T^{1+2\nu}}\right).
\end{equation}
Furthermore, we expect that near the $T=0$ transition, long time
dynamics will be governed by activation over barriers. Hence we expect
\begin{equation}
T^{1+\nu}\rho_{\rm lin}=\frac{1}{\tau}=A\exp(-\Delta E(T)/T),
\end{equation}
where $\Delta E$ is the typical barrier that a vortex has to
cross to move a distance $\xi$.
One can define a  barrier height exponent $\psi$ by
$\Delta E\sim \xi^{\psi}\sim T^{-\psi\nu}$ in terms of which
\begin{equation}
T^{1+\nu}\rho_{\rm lin}=A\exp(-C/T^{1+\psi\nu}).
\label{arrhenius}
\end{equation}
We are able to obtain a rough estimate for $\psi$ from
our data of the linear resistivity.

In a finite system, the $I$-$V$ characteristics will also
depend on the ratio $L/\xi$. One can generalize
the scaling function, Eq.~(\ref{scale_iv}), to account for finite
size effects as follows:
\begin{equation}
\frac{E}{J\rho_{\rm lin}} = \tilde{g}\left(
\frac{J}{T^{1+2\nu}},L^{1/\nu}T
\right).
\label{iv-scale}
\end{equation}
Now we are left with a rather complicated scaling function
since it depends on two variables. To simplify the analysis we first
estimate $\nu$ by determining the
current where $E / (J\rho_{\rm lin}) = 2$, at which point non-linear
effects start to become significant. Denoting these values
of $J$ by $J_{\rm NL}$, then,
from Eq.~(\ref{iv-scale}), it follows that
\begin{equation}
{J_{\rm NL} \over T^{1+2\nu} } = \tilde{\tilde{g}} \left(L^{1/\nu}T\right) ,
\label{scale_jnl}
\end{equation}
where $\tilde{\tilde{g}}$ is another function. Hence we determine
$\nu$ by requiring that the scaling in Eq.~(\ref{scale_jnl}) is satisfied. 
We then collect data for sizes and temperatures such that $L^{1/\nu}T$
is constant. The scaling function in Eq.~(\ref{iv-scale}) then
only depends on {\em one} variable, and so data 
for $E/J\rho_{\rm lin}$ for different sizes should scale when plotted against 
$J/T^{1+2\nu}$, with the {\em same} value of $\nu$ as obtained from the scaling
of $J_{\rm NL}$.
We find, in fact, that the results are
only weakly dependent on the second argument of Eq.~(\ref{iv-scale}).

\section{Results for the gauge glass}
\label{gg}

In this section we consider the critical behavior of the gauge
glass model with and without screening. Recall that the distribution of the
(shifted) fluxes is given by Eq.~(\ref{dist-gg}).

\begin{figure}[bt]
\centerline{\epsfxsize 7cm \epsfbox{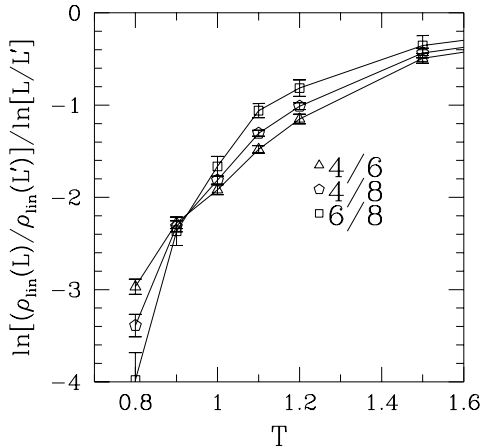}}
\caption{Plot of 
$\ln[\rho_{\rm lin}(L)/\rho_{\rm lin}(L^{\prime})]/\ln[L/L^{\prime}]$
versus $T$ for the gauge glass with $\lambda_0\to \infty$. The curves 
intersect at $T_c=0.93\pm 0.05$. At the intersection point, the $y$--value
is approximately $-2.2$, corresponding to $z_{\rm GG}\approx 3.2$.}
\label{gg_l_int}
\end{figure}

\subsection{No screening, $\lambda_0\to \infty$}

\begin{figure}[bt]
\centerline{\epsfxsize 7cm \epsfbox{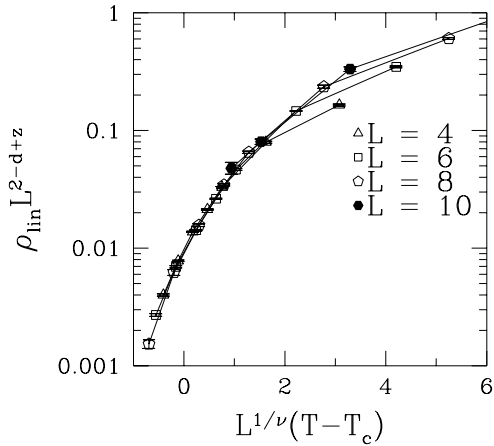}}
\caption{Scaling plot of the lineage resistivity for the gauge glass
with $\lambda_0\to \infty$. Using $T_c=0.93$ and $z_{\rm GG}=3$ from 
Fig.~\protect\ref{gg_l_int} we obtain $\nu_{\rm GG}=1.3 \pm 0.3$.}
\label{gg_l_sc}
\end{figure}

For the gauge glass with no screening we have measured the linear
resistivity $\rho_{\rm lin}$ as a function of temperature.
In Fig.~\ref{gg_l_int}
we show data of $\rho_{\rm lin}$ plotted according to the
intersection method vs. $T$ for sizes $L=4,6,8$. We were not able to include
data from $L=10$ into this plot since we could not equilibrate the
systems down to the lowest temperatures ($T=0.8,0.9$).
All curves intersect at about $T=0.93 \pm 0.05$ indicating a phase
transition to a vortex glass. The corresponding $y$--axis value 
at the intersection point is $1-z\approx -2.2$, therefore 
$z_{\rm GG}\approx 3.2$. Having established these values, we tried a scaling 
plot according to Eq.~(\ref{rho_lin_scale}) and the result is shown
in Fig.~\ref{gg_l_sc}. Best scaling was achieved with $T_c=0.93$,
$z_{\rm GG}=3$ and $\nu_{\rm GG}=1.3 \pm 0.3$. 
Only far away from the transition point does one observe deviations 
from scaling, which is expected for such small sizes and high
temperatures, but the overall scaling works quite well.

It is interesting to compare this result with earlier Monte Carlo
simulations of the gauge glass without screening in the phase
representation by Reger et al.\cite{reger91} These authors
did a finite size scaling analysis of static quantities which indicated a
finite temperature transition, but they could not
completely rule out the possibility
that the lower critical dimension is $d_l \simeq 3$.
The clear intersection of the data in Fig.~\ref{gg_l_int},
however, strongly confirms the notion that there is a finite temperature
transition in the three--dimensional gauge glass model, and hence $d_l<3$.
Additionally, our correlation length exponent $\nu_{\rm GG}$
agrees well with the 
estimate given by Reger et al.\cite{reger91},
$\nu_{\rm GG}=1.3 \pm 0.4$. There is, however, a
considerable difference between our estimate of the dynamic critical
exponent $z\approx 3.1$ and theirs,
$z=4.7 \pm 0.7$. It is possible, though, that the {\em dynamical} universality
classes of the models in the phase and vortex representations may be different,
even though the static behavior is the same. If so, there is no contradiction
in the results.

\subsection{Strong Screening, $\lambda_0\to 0$} 
\label{gg_l0}

In this paragraph we review quickly the results for the gauge glass model 
with strong screening found earlier by WY,
in order to compare them in the next section with our
data for the $d=3$ XY--spin glass model with screening.
As shown by WY the vortex 
glass transition in the gauge glass
is destroyed by screening of the vortex--interactions.
The main indication for the lack of a transition
at finite $T$ was the absence of an intersection if the resistivity was
plotted according to the intersection method. A scaling plot of the
current--voltage characteristics for different temperatures and 
sizes also revealed $T_c=0$ and $\nu_{\rm GG}=1.05\pm 0.1$. Finally,
the barrier exponent $\psi$, as defined in Eq.~(\ref{arrhenius}),
was determined to be close to zero, so 
the conclusion was drawn that energy barriers diverge only weakly, possibly 
logarithmically, as one approaches the zero temperature transition.

\begin{figure}[t]
\centerline{\epsfxsize 7cm \epsfbox{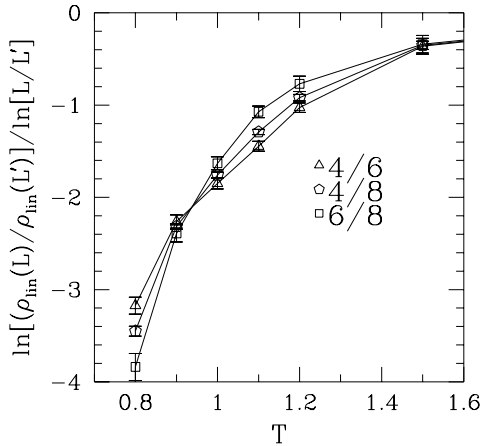}}
\caption{Plot of
$\ln[\rho_{\rm lin}(L)/\rho_{\rm lin}(L^{\prime})]/\ln[L/L^{\prime}]$
versus $T$ for the $\pm J$ XY--spin glass and $\lambda_0\to \infty$. 
The curves intersect at
$T_c=0.94\pm 0.05$. At the intersection point, the $y$--value
is approximately -2.2, corresponding to $z_{\rm CG}\simeq 3.2$.}
\label{xy_l_int}
\end{figure}

\section{Results for the $\pm J$ XY--spin glass} \label{xy}

Recall that the only difference between the $\pm J$ XY--spin glass and the
gauge glass discussed in the last section is that the distribution of shifted
fluxes is given by Eq.~(\ref{dist-sg}) rather than by Eq.~(\ref{dist-gg}).

\subsection{No screening, $\lambda_0\to\infty$}

As already discussed in the introduction, the $\pm J$ XY--spin glass
is known to have no finite--temperature transition to an
ordered state below four dimensions\cite{jain86}. For the $d=3, \pm J$ model
one can, however, identify a chiral glass transition in Monte Carlo
simulations due to freezing out of the discrete degrees of freedom,
as has been done by Kawamura et al.\cite{kawamura95/6} 
The associated chiral glass exponents estimated
in the phase representation with periodic boundary
conditions\cite{kawamura95/6} 
are $\nu_{\rm CG}=1.5 
\pm 0.3$ and $\eta_{\rm CG}=-0.4 \pm 0.2$. Subsequent work with free
boundary conditions\cite{kawamura97} finds similar values, $\nu_{\rm CG}=1.3 
\pm 0.2$ and $\eta_{\rm CG}=-0.2 \pm 0.2$.

\begin{figure}
\centerline{\epsfxsize 7cm \epsfbox{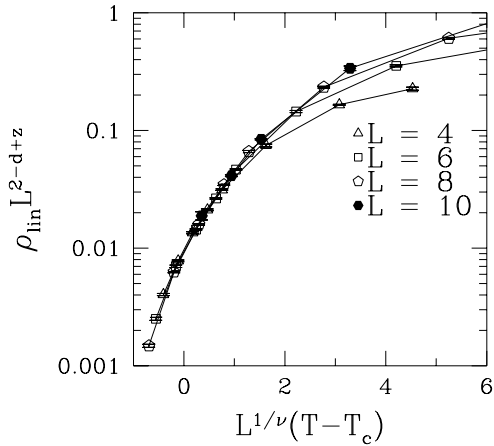}}
\caption{Scaling plot of the lineage resistivity for the $\pm J$
XY--spin glass
with $\lambda_0\to \infty$. Using $T_c=0.94$ and $z_{\rm CG}=3.1$ from
Fig.~(\protect\ref{xy_l_int}) we obtain $\nu_{\rm CG}=1.3 \pm 0.3$.}
\label{xy_l_sc}
\end{figure}

Figure \ref{xy_l_int} displays a plot of our data for
$\rho_{\rm lin}$ according to
the intersection method vs. $T$ for the XY--spin glass. One observes, very
similarly to Fig. \ref{gg_l_int}, an intersection point at $T=0.93 \pm
0.05$ and a dynamic critical exponent $z_{\rm CG}\approx 3.2$. Also, 
the scaling
plot of $\rho_{\rm lin}$ shows best results with almost the same values
as in the long range gauge glass case, namely $T_c=0.93$, $z_{\rm CG}=3.1$ and
$\nu_{\rm CG}=1.3 \pm 0.3$.
This result indicates a
finite--temperature transition into a chiral glass state for the $d=3$
XY--spin glass and thereby confirms Monte Carlo results performed in 
the phase representation.\cite{kawamura95/6}
Very surprisingly, we find that
our data for the linear resistivity is virtually indistinguishable
from the corresponding measurements of the
gauge glass model. We observe a maximum deviations of $1.5\sigma$. We will come
back to this in the last section.

\subsection{Strong screening, $\lambda_0\to 0$} \label{xy_s}

\begin{figure}
\centerline{\epsfxsize 7cm \epsfbox{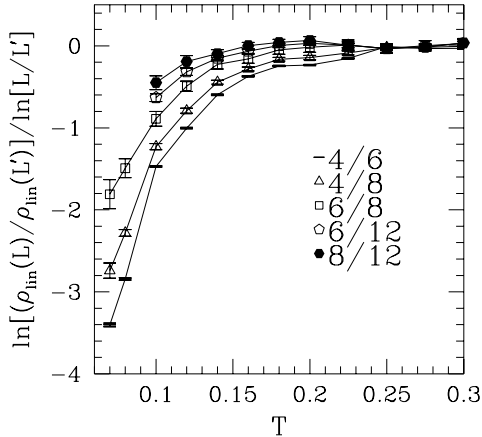}}
\caption{Plot of
$\ln[\rho_{\rm lin}(L)/\rho_{\rm lin}(L^{\prime})]/\ln[L/L^{\prime}]$
vs. $T$ for the XY--spin glass model with $\lambda_0\to 0$.
In contrast to Fig.~\protect\ref{xy_l_int} there is no intersection over
the entire temperature range, indicating the absence of a phase
transition into a chiral glass.}
\label{xy_s_int}
\end{figure}

In Fig.~\ref{xy_s_int} we show the linear resistivity plotted according
to the intersection method vs. $T$. On can see, that there is no
apparent intersection over the entire temperature range that we have
been able to simulate, i.~e., down to $T=0.07 $ for $L\le 8$ and $T=0.1$
for $L\le 12$. At high temperatures all curves merge,
since the correlation length becomes shorter then the system size and
the data of $\rho_{\rm lin}$ for different sizes are the same. 
This rules out a transition down to 1/5
of the critical temperature of
the system without disorder, $T_c=0.331$ (see WY), and therefore
strongly suggests the absence of a chiral glass transition at finite
temperature, in agreement with work by Kawamura and 
Li.\cite{kawamura97}

\begin{figure}
\centerline{\epsfxsize 7cm \epsfbox{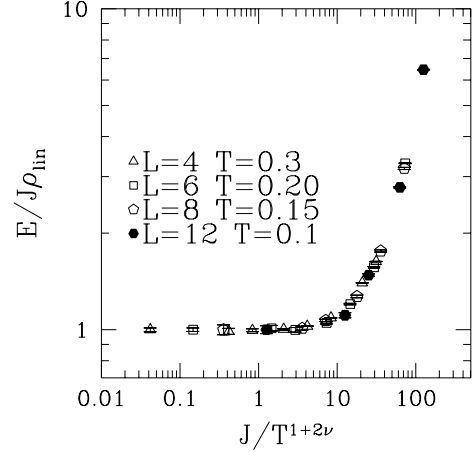}}
\caption{Scaling plot of the $I$--$V$ characteristics with $T_c=0$ and
$\nu_{\rm CG}=1.05 \pm 0.1$, according to Eq.~(\protect\ref{iv-scale}), 
choosing sizes and temperatures
such that $L^{1/\nu_{\rm CG}}T$ is roughly constant.}
\label{xy_iv}
\end{figure}

Next we studied the current--voltage characteristics of our model in
order to determine $\nu$. Figure~\ref{xy_iv} shows a scaling plot of
different $I$--$V$ curves according to Eq.~(\ref{iv-scale}). From the scaling
of the nonlinear current $J_{\rm NL}$ 
we estimated $\nu_{\rm CG}=1$, and then chose
sizes and temperatures for the data in Fig.~\ref{xy_iv} such that
$L^{1/\nu_{\rm CG}}T=const.$, and hence the 
second argument in Eq.~(\ref{iv-scale}),
remained roughly constant. The data is seen to scale very well with
$T_c=0$ and $\nu_{\rm CG}=1.05 \pm 0.1$. We
also attempted scaling our data with
an appropriate scaling function for finite $T_c$, and found that
scaling works only moderately well with $T_c=0.04$ and $\nu_{\rm CG}=1.05$. 
We, therefore, conclude that the transition is very likely to occur at
$T_c=0$, but we cannot completely rule out a finite, though extremely
small $T_c$. 

We also determined the barrier exponent $\psi$ by plotting
$T^2\rho_{\rm lin}$ over $1/T$ as was done in Fig.~4 of WY for the gauge
glass. The data
for $L=12$ follows almost a straight line indicating Arrhenius behavior
and therefore $\psi\simeq 0$. As in the gauge glass case one has to be
careful though, since such an estimate does not allow for finite--size 
corrections and is only observed over a small range of temperatures.
It is also possible, that we only measure an effective exponent and the
true value of $\psi$ changes as one gets closer to $T=0$. In any
case, $\psi \simeq 0$ would suggest, that barriers increase only very slowly,
possibly
logarithmically, as one approaches the zero temperature chiral glass
transition.

Again it is interesting to compare these results with those obtained by
WY for the gauge glass with screening: they agree perfectly with in the
errors, namely $T_c=0$, $\nu_{\rm GG}=1.05$ and $\psi\simeq 0$, as described in
Sec.~\ref{gg_l0}.
Not only do the final estimates of the exponents for the gauge glass and $\pm
J$ XY--spin glass agree but also, as in
the case without screening, 
the individual numerical values of $\rho_{\rm
lin}$ and data from the $I$--$V$ characteristics all agree within the 
errorbars, the maximum discrepancy being $1.5\sigma$.

\section{Summary and Discussion} \label{conclude}

In this article we have presented a Monte Carlo study of the gauge glass
model and the chiral glass transition in the
XY--spin glass model with and without screening, in
the vortex representation. We have computed dynamic quantities such as
the linear resistivity and current--voltage characteristics and used
finite--size scaling techniques to extract the critical behavior of
these models.

Our main results are:
\begin{enumerate}
\item
In the absence of screening there is 
a finite temperature transition in both cases with
numerically indistinguishable exponents given in Table. 1.
\item
In the presence of strong screening, there is a transition
at zero temperature in both cases.
The correlation length exponent is the same for the two models,
as shown in Table 1.
\item
Not only do the gauge glass transition and the chiral transition in the
XY--spin glass model appear numerically 
to be in the same universality class, but even
the individual data points for the 
current--voltage characteristics are virtually indistinguishable.
\end{enumerate}

\renewcommand{\arraystretch}{1.25}
\begin{table}
\label{result}
\begin{tabular}{|c|d|c|d|d|}
Screening & $T_c$ & $\nu$ & $z$ & $\psi$ \\
\hline\hline
$\lambda_0 \to \infty$ & 0.94 $\pm$ 0.05 & 1.3 $\pm$ 0.3 &
$\approx$ 3.1 & n/a \\
$\lambda_0 \to 0$ & $0$ & 1.05 $\pm$ 0.1 & $\infty$ & $\simeq 0$\\
\end{tabular}

\vspace{0.3cm}
\caption
{
Critical temperatures and exponents of the presumed common
universality class of the gauge glass and the chiral glass in $d=3$.
$\nu$ is the correlation length exponent, $z$ is the dynamical exponent, and
$\psi$ is the barrier exponent for the $T=0$ transition in the strong screening
limit.
}
\end{table}

Earlier work which provided evidence for a finite temperature chiral glass
transition\cite{kawamura95/6,kawamura97}
used the phase representation and constructed the chiralities
(vortices) indirectly
from the spin configurations. This is only sensible at moderate to
low temperatures where correlations in the angles of nearest neighbor spins
become significant. Our work is the first which demonstrates the existence of
a chiral glass transition using the vortex representation.
In our model
vortices are well defined at {\em all} temperatures and so we expect that
the region over which scaling behavior is obtained will be larger than in the
earlier work in the phase representation. We therefore feel that our results
make the existence of the chiral glass transition more convincing.

There is also support in two dimensions for the idea that the chiral glass and
gauge glass transitions are in the same universality class since in
both cases one finds\cite{kawamura91,ray92,hyman95,bokil95,bokil96}
$T_c=0$ and $\nu\approx 2$.

However it is unclear to us theoretically
why the gauge glass and the chiral glass transition in
the $\pm J$ XY--spin glass {\em should} be
in the same universality class. For the XY--spin glass, the important low
energy states are those where $n^\alpha_i - b^\alpha_i = 0$ on links where
$b^\alpha_i = 0$ (corresponding to an unfrustrated square on the original
lattice) and $n^\alpha_i - b^\alpha_i = \pm 1/2$ on links where $b^\alpha_i =
1/2$. Thus, as first noted by Villain,\cite{villain77} one has a random Ising
model with long range antiferromagnetic interactions,
\begin{equation}
{\cal H} =  -{1 \over 4}
\sum_{\langle i,j\rangle } G(i-j) \epsilon_i \epsilon_j S_i S_j ,
\label{ham-xy}
\end{equation}
where the $\epsilon_i$ are quenched variables taking values 0 or 1, and the
$S_i$ are statistical Ising-like variables which take values $\pm 1$. For the
gauge glass one cannot make an analogous transformation and Eq.~(\ref{hv})
corresponds to an Ising model {\em in a random field}, which is not expected to
be in the same universality class as Eq.~(\ref{ham-xy}).
We do not, therefore, understand why the numerical values of the exponents are
the same within the uncertainties. Even more surprising is that the individual
I--V values for the two models are virtually indistinguishable. We would expect
there to be a more clearly visible difference in other properties. Perhaps for
some reason, the random field aspect of the gauge glass is irrelevant, or
perhaps the critical behaviors of the two models just happen by coincidence
to be very close. It would be interesting to check our results by studying both
models by
alternative techniques such as domain wall renormalization group
methods.

The correlation length exponent for the unscreened models is also very similar
to that of the Ising spin glass\cite{kawashima95} with short range interactions.
Again, it is not obvious to us why this
should be the case. While the model in Eq.~(\ref{ham-xy}) has Ising variables,
and the ingredients of randomness and frustration necessary for a spin glass,
it also has long range interactions, unlike the Ising spin glass.

Finally, it is noteworthy, that 
earlier results for the gauge glass\cite{bokil95} indicated that
the universality class changes (and hence $T_c$ becomes zero)
for any non-infinite value of the bare screening length.
By contrast, Kawamura and Li\cite{kawamura97} have
argued that the transition in the
chiral glass persists down to a finite value of $\lambda_0$.
It would, therefore, also be interesting to study these models 
with an intermediate range of screening in the vortex representation.

\begin{acknowledgments}
We wish to thank Hemant Bokil, Muriel Ney--Nifle and Christian Pich
for useful discussions.
This work has been supported by NSF grant DMR 94--11964. The work of 
CW has also been supported in part by a fellowship of the German Academic
Exchange Service (Doktorandenstipendium HSP II/AUFE).
We would like to thank the Maui High Performance Computing Center for 
an allocation of computer time.
\end{acknowledgments}


\begin{thebibliography}{99}
\bibitem{comment_z}
  Note also that without disorder
  (i.~e., all ${\bf b}_i=0$) this model is just a dual representation
  of the $XY$--model\protect\cite{dasgupta81,kleinert}
  (with the Villain potential), in which the temperature scale
  has been inverted.
\end{thebibliography}
\end{document}